# When do Default Nudges Work?

Carl Bonander, Mats Ekman, and Niklas Jakobsson*

August 2023

Abstract

Nudging is a burgeoning topic in science and in policy, but evidence on the effectiveness of nudges among differentially-incentivized groups is lacking. This paper exploits regional variations in the roll-out of the Covid-19 vaccine in Sweden to examine the effect of a nudge on groups whose intrinsic incentives are different: 16-17-year-olds, for whom Covid-19 is not dangerous, and 50-59-year-olds, who face a substantial risk of death or severe disease. The response is strong in the younger but absent in the older age group, consistent with the theory that nudges work best for choices that are not meaningful to the individual.

*Bonander: University of Gothenburg, School of Public Health and Community Medicine, Institute of Medicine, 405 30 Gothenburg, Sweden (email: carl.bonander@gu.se); Ekman: Karlstad University, Karlstad Business School, 651 88 Karlstad, Sweden (e-mail: mats.ekman@kau.se); Jakobsson (corresponding author): Karlstad University, Karlstad Business School, 651 88 Karlstad, Sweden (e-mail: niklas.jakobsson@kau.se). We thank Siri Jakobsson Støre, Andreas Kotsadam, Katarina Nordblom, and participants from various seminars and conferences for providing valuable comments and suggestions. The authors have no financial or other material interests related to this research to disclose. We acknowledge funding from Jan Wallanders och Tom Hedelius stiftelse samt Tore Browaldhs stiftelse (grant number P22-0018), as well as from the Swedish Research Council for Health, Working Life, and Welfare (Forte; grant number 2020-00962).

# 1 Introduction

Nudges have established themselves as attractive policy tools for influencing actions without removing options. Society sometimes benefits from people taking certain decisions, but individuals may lack incentives to take them, or they may prefer to make socially harmful choices. Numerous instances have been found where small changes in the choice architecture bring vast changes in decisions. These range from situations involving participation in retirement-savings plans (Atalay *et al.*, 2014; Madrian & Shea, 2001) to eating healthy (Cheung *et al.*, 2019) or eliciting organ donations (Li *et al.*, 2013). Despite this burgeoning interest, a largely unresolved empirical issue is whether nudges sway those benefiting the most from pension savings, eating healthier food, etc., or those who benefit less (de Ridder *et al.*, 2021; Szaszi *et al.*, 2018).

There are two extreme and opposing interpretations of the empirical literature on nudges. In one, individuals lack volition, so the tiniest nudge may cause individuals to choose something other than the option that they actually prefer. In the other, individuals have no reason to prefer one option over another, and therefore simply pick the default (*cf.* de Ridder *et al.*, 2021). Comparing people with different intrinsic motivations to select a particular option while holding the choice situation constant is one way to find out which interpretation is most accurate. The nudge could be tested on two groups, each with its own control, and each with different intrinsic incentives to choose the nudge option. If both nudged groups differ from their controls, volition is weak. If the intrinsically motivated group makes the same choice with or without a nudge, volition is strong. This paper does precisely this by examining regional differences in the roll-out of the Covid-19 vaccine in Sweden. The Region of Uppsala was unique among Sweden's 21 regions in that its authorities summoned a large share of the Region's inhabitants to pre-booked appointments, rather than simply opening up telephone lines and websites for people to make arrangements on their own. Because anyone who receives a pre-booked appointment could ignore it and book an appointment (or not) as in the other 20 regions, this is a nudge. In this way, the authorities changed the default option to a suggested vaccination time compared to no suggested time.

For our purposes, the crucial aspect of these pre-booked appointments is that they were sent to individuals aged 50 or over and to individuals between the ages of 16 and 17 (except for immunocompromised individuals and hospital staff, pre-booked appointments were not made for those aged between 18 and 49). Given the vast and well-known age gradient in the risk of death or severe disease from Covid-19 and the established connection between risk perception and protection behavior (Brewer *et al.*, 2007), the pre-booked vaccinations in Uppsala provide an excellent opportunity to study the effectiveness of default nudges in age-groups that can be expected to be affected differently by the nudge. Our main empirical strategy is to compare the Region of Uppsala and a synthetic control region. We estimate the impact of the roll-out of vaccines on vaccination rates for persons aged between 50 and 59 and persons aged between 16 and 17, finding the nudge to have no robust effect on the older group but a vast and robust effect on the younger one.

Our results are consistent with the theory that nudges work best when the nudged individual has little reason to prefer one option over another. Some theoretical analyses incorporate intrinsic motivation into decision problems where the agent is sometimes nudgeable and sometimes not. Löfgren and Nordblom (2020) develop a model in which individuals who face an unimportant decision are more likely to take it inattentively and so are more nudgeable the less important they perceive the decision to be. A contrasting approach is taken by de Ridder *et al.* (2021), who piece together general propositions from empirical regularities. Like Löfgren and Nordblom, they stress the personal preferences of the nudged, arguing that one cannot be nudged into something one does not want, and that nudges are likely most effective for those who are indifferent. In contrast to Löfgren and Nordblom, they place less emphasis on the relevance of nudges to "System 1" cognitive processes, i.e., the fast and unconscious response of the brain (Kahneman, 2011), arguing that deliberation ("System 2") might lower nudgeability, but not make it impossible. Our findings are consistent with Löfgren and Nordblom (2012) and de Ridder *et al.* (2021). However, it may also be that older individuals are more attentive for reasons that are irrelevant to the perceived importance of getting vaccinated against Covid-19. At any rate, the medical importance increases with age.

Another perspective is provided by Alcott and Kessler (2019), who develop a model that allows moral prices and moral utility (e.g., others' expectations) to influence decisions, in addition to inattention. These aspects may be related to perceptions of risk and age to the extent that individuals believe that getting vaccinated will stop transmission of the virus. Thus, there may be little personal gain but much pressure from others to get vaccinated for the young, and more personal gain for the old. Unfortunately, we cannot assess the relative importance of attention, importance, moral prices and moral utility in our data, and we must therefore remain silent on the exact reason for the lack of nudgeability of the 50-59-year-olds compared to younger persons.

Several findings of null effects of nudges can be explained with reference to incentives, skills, or expertise of the nudged. In this vein, Löfgren *et al.* (2012) find no effect of pre-set defaults on decisions to offset carbon emissions for air travel among environmental economists at a conference (i.e., people with expertise on the issue of the nudge). Similarly, in influencing debtors to repay their debts and school children to apply for higher education, Holzmeister *et al.* (2022) and Ilie *et al.* (2022), respectively, find no effect of nudging messages, plausibly because the incentives are already clear to the recipients. Finally, in Bonander *et al.* (2022) we examine Region Uppsala's vaccination nudge on 16-17-year-olds using partly the same data as in the present paper, finding a large effect but noting the little personal concern of the nudged individuals. This paper extends our previous analysis by contrasting these results with a group with high personal concern, arguably allowing us to study the moderating effects of intrinsic motivation.

Despite receiving widespread attention in the literature, we are not aware of any previous attempts to estimate the impact of nudges on two groups with different intrinsic motivations. The study that comes closest to doing this is the retirement-savings study by Madrian and Shea (2001), who find that the effect of the pre-set default is lower (though still positive) for older persons. Age could imply a differential in intrinsic motivation because older employees are generally closer to retirement. In an extensive literature review, Szaszi *et al.* (2018) note how several factors across different settings can moderate the effects of nudges and highlight the importance of quantifying effect heterogeneity in future studies.

Our setting has at least two advantages for measuring the impact of a nudge and how it differs across groups with different intrinsic motivations. Firstly, the speed with which the vaccine was rolled out meant that the letters with pre-booked appointments were sent to both the quinquagenarians and the teenagers at roughly similar times, on 5$^{\text{th}}$ May of 2021 for the 50-59-year-olds and on 15$^{\text{th}}$ July for the 16-17-year-olds. This time of the year is off-season for Covid-19 in Sweden and there was reason to believe it to be so from experiences in the previous year, meaning that the two groups made their vaccination decisions in roughly the same pathogenic environment. In early May, there was more virus in circulation than in July, but at the time of the vaccine roll-out, there was talk of how the vaccine would require up to two weeks' time before it would fully protect the recipients. Thus, allowing some time for the vaccine to take effect, both groups could expect to be effectively vaccinated in the same low-risk environment.

Secondly, the age-stratified mortality and morbidity of Covid-19 create a natural differentiation in intrinsic motivation for the two groups to get vaccinated. In one estimate, pre-vaccine seroprevalence data indicate an over 400-fold difference in the infection fatality rate between persons aged 19 or younger and those aged 50 to 59 (Pezzullo *et al.*, 2023). Other estimates are less dramatic but nevertheless suggest a 100-fold difference (O'Driscoll *et al.*, 2021; Ghisolfi *et al.*, 2020). In any event, the risk difference between 50-59-year-olds and 16-17-year-olds is clear. Thus, we have a clean setting in which nature has differentiated intrinsic motivation, and the authorities have acted with sufficient celerity to ensure decision-makers in our study make their choices in roughly the same environment.

The remainder of the paper is organized as follows. Section 2 describes the Swedish case and vaccination practices in Swedish regions. Section 3 presents data and methods used for estimating the effects of the nudge. Section 4 presents the results of the empirical analysis as well as several robustness checks. Section 5 compares the findings with earlier studies and discusses the implications for policies related to default nudges.

## 2 Vaccine Nudges in Swedish Regions

We contacted all 21 regions of Sweden to ask if the regional health authorities deviated from the standard practice of opening up telephone lines and websites to let people make their own vaccination arrangements. It was not standard practice to send out letters with pre-booked appointments for vaccinations against Covid-19 in Sweden during the vaccine roll-out. However, for particular age groups, authorities in a number of regions made special arrangements. The health authorities in the Region of Uppsala were the only ones to send letters with pre-booked appointments to everyone aged 16-17 and 50-59 years. Except for immunocompromised individuals and severely handicapped individuals who rely on assistance for almost everything, everyone up to the age of 50 made their own appointments in all of Sweden, including Uppsala. The health authorities in a total of five other regions made use of nudges in some way, but in every case, the use was much more limited compared to the Region of Uppsala.

The regions that come closest to using nudges in the same way as Uppsala are those of (1) Dalarna, where the health authorities sent letters with pre-booked appointments to persons aged 80 or over, as well as to particular high-risk individuals of all ages, and (2) Västmanland, where individuals aged 75 or above were sent letters with pre-booked appointments. In (3) the Region of Jönköping, the health authorities made no use of nudges before 6$^{\text{th}}$ September 2021, but then sent letters with pre-booked appointments to "severely immunocompromised individuals" (their words used in email correspondence with us) for a booster.

No other regions used pre-booked appointments. However, in (4) the Region of Västra Götaland, persons aged 85 or above were sent postcards informing them that they would receive a telephone call helping them to book an appointment. In addition, everyone born in 1946 or before (i.e., people at least 75 years old) was contacted by their local health providers so that they could find a suitable time for a vaccination appointment. In this region, immunocompromised individuals of all ages above 18 were also contacted to set vaccination appointments. This is a nudge in the sense that it changes the choice architecture and the contacted individuals could still say no or hang up and then make their own arrangements. In a similar vein, the health authorities of (5) the Region of Jämtland-Härjedalen contacted all 80-year-olds and over by telephone, but otherwise did not use any nudges.

## 3 Empirical Methodology

### 3.1 Data

To estimate the effect on vaccinations of the letters with pre-booked vaccination appointments, we use weekly panel data on the share of vaccinated individuals in the 21 Swedish regions (NUTS3) obtained from the Public Health Agency of Sweden (2021). We have previously used some of this data to study the treatment effect on 16-17-year-olds only (Bonander *et al.*, 2022). Our data contain the share of vaccinated individuals in the age groups 16–17, 18–29, 40–49, and 50–59 from weeks 1 to 46 in 2021. We choose the first week in 2021 as the starting point for our analysis since this includes the first vaccinations in the studied age groups. We choose week 46 as the end date because that was when the Swedish government announced mandatory vaccinations for participation in certain public events. Periods in which fewer than four individuals were vaccinated are treated as zero for reasons of personal integrity. Thus, some data are missing in weeks when very few individuals got vaccinated. However, this is mainly a problem in the early stages of the vaccine roll-out, when mostly high-risk individuals (e.g., chronic lung disease, cancer, and diabetes) were vaccinated. For descriptive purposes, we also have data on the share vaccinated in treated and nearby municipalities in week 49.

We include several covariates that may be relevant confounders following recent empirical findings on relevant predictors of attitudes towards COVID-19 vaccination (Bonander *et al.*, 2022): the share of foreign-born individuals living in the region and the share of the population with at least three years of higher education in 2020 (Statistics Sweden, 2021); the share of COVID-19 deaths in 2020 (Public Health Agency of Sweden, 2021); the share of the population that has received financial aid at any point during the past year in 2020; the number of adolescents 10-24 per 100000 inhabitants who have received care due to alcohol addiction in 2019; the share of the population that has access to a fast broadband connection (100 Mbit/s) in 2020; the share of the population with high trust in how the Swedish health care system handled the COVID-19 pandemic in 2020; and NEETs (the share of 17-24-year-olds that neither study nor work) in 2020 (Kolada, 2021). These variables do not necessarily have a direct bearing on attitudes towards vaccinations, but they are plausible measures of general satisfaction with authorities including health authorities, and attitudes to health more broadly.

## 3.2 The Synthetic Control Method

We use the synthetic control method to construct a synthetic control that closely resembles Uppsala in terms of pre-intervention characteristics, from a weighted combination of all other Swedish regions (Abadie et al., 2010; Abadie, 2021). The synthetic control method is designed to estimate the impact of policy interventions affecting one unit (e.g., country, region, or municipality) when a small number of control units are available. It is a data-driven approach for estimating counterfactuals—i.e., what would have happened without the nudge—which determines the weighted combination of untreated regions that provides the closest match to the treated region with regard to pre-intervention outcomes and covariates. The weighted average vaccination uptake from the synthetic control group then provides the counterfactual trend of the vaccination share for the treated region; i.e., it predicts how the vaccination rates would have turned out in the absence of the nudging intervention. For a detailed presentation of the method, see Abadie (2021). Following Abadie and Gardeazabal (2003), we employ a nested optimization routine to determine the optimal set of unit weights (W), which determine each untreated unit's contribution to a synthetic control, and variable importance weights (V), which determine the likelihood of a good match on strong predictor outcomes.

Specifically, let $J + 1$ be the number of regions indexed by $j$, and let $j = 1$ denote Uppsala. The regions in the sample are observed for periods $t = 1, 2, ..., T$, where $T_0$ is the number of pretreatment periods. Next, we define two potential outcomes: $Y_{jt}^I$ is the potential outcome if region $j$ is exposed to the nudge intervention in time $t$, and $Y_{jt}^N$ is the corresponding potential outcome without the nudge intervention, which is unobserved in Uppsala after $T_0$. The goal of the analysis is to measure the post-treatment effect in Uppsala (region $j = 1$), defined as $\alpha_{1t} = Y_{1t}^I - Y_{1t}^N$. Since $Y_{1t}^N$ is unobserved, we use the synthetic control method to impute it. The synthetic region is constructed as a weighted average of the control regions $j = 2, ..., J + 1$ from the donor pool of control regions and represented by a vector of weights $W = (w_2, ..., w_{J+1})'$ with $0 \leq w_j \leq 1$ and $w_2 + \cdots + w_{J+1} = 1$. Each choice of W gives a particular set of weights and characterizes a possible synthetic control. For each age group, we choose the W that minimizes the pre-intervention mean squared error between Uppsala and its synthetic control on the eight predictors of the outcome variable presented in the data description and the outcome variable itself. The predictors are recorded once at the end of the year 2020, or 2019 (depending on data availability). Since there are zero vaccinated 16-17-year-olds in the weeks before intervention we match on the share of vaccinated 18-29-year-olds in the analyses of the younger age group. In the main analyses we use the average vaccination shares in the weeks before the intervention, but we test 14 different ways of including lagged values of the outcome variable, as suggested by Ferman *et al.* (2020). To assess if the age-specific estimates differ from each other, we also assess the difference in effect estimates between the two age groups in the treated region by week relative to the start of the vaccinations.

# 4 Results

## 4.1 Descriptive municipal comparison

As a first step, we compare the vaccination shares in treated municipalities to neighboring municipalities. Figure 1 shows vaccination shares for 16-17-year-olds (left) and 50-59-year-olds (right) in treated municipalities (in region Uppsala) and nearby (non-treated) municipalities. The difference in vaccination shares between treated and untreated municipalities is bigger for the younger than for the older age group. A comparison between the eight treated municipalities and their eight bordering neighbors shows that the vaccination rate was 12.9 percentage points (95% CI: 9.0, 16.8) for the younger age group, and 3.6 percentage points for the older age group (95% CI: 1.4, 5.9). This descriptive analysis indicates that there may be an effect of the vaccination nudge, especially in the younger age group.

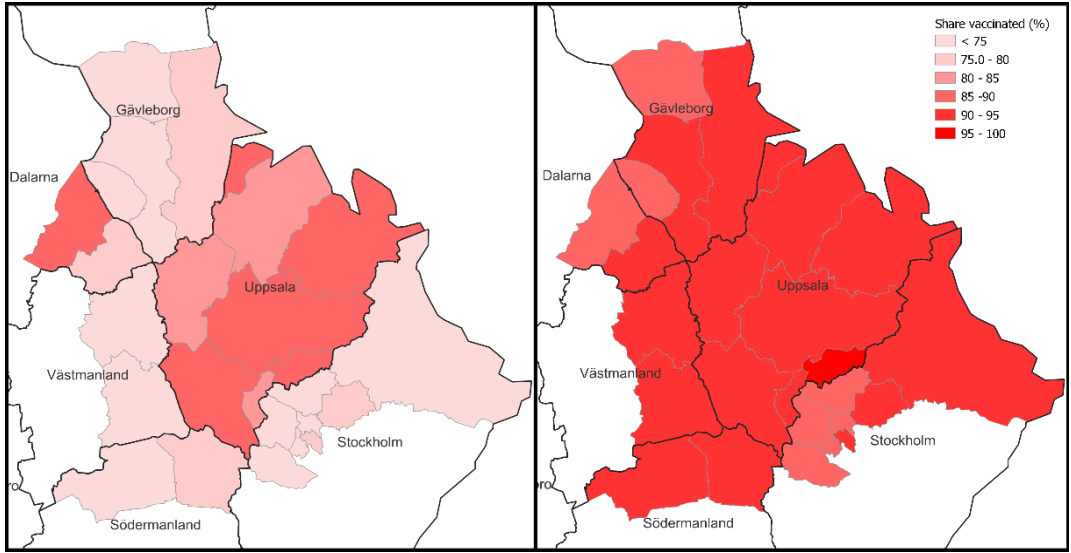

Figure 1. Share vaccinated 16-17-year-olds (left) and 50-59-year-olds (right) in treated municipalities (Uppsala) and nearby (non-treated) municipalities, week 49

## 4.2 Regional analyses

Figure 2 shows the trajectories of the share of vaccinated 16–17-year-olds (left panel) and 50–59-year-olds (right panel) in the treated region (Uppsala) and the average of all other Swedish regions during the first 46 weeks of the year 2021. The vaccination rates follow similar paths at the beginning of the year. However, the trajectories diverge after week 30 for the 16–17-year-olds and from week 16 for the 50–59-year-olds.

Figure 3 shows that, prior to treatment, the trajectories of vaccinations in the treated region and its synthetic counterpart are very similar in both age groups. Furthermore, Table 1 compares the values of key predictors for the treated region before treatment, with the same values for the synthetic region and the average of all 20 control regions in the time periods before treatment. For all predictors, except share with financial aid, the treated region and its synthetic counterpart have a better fit compared to the average. The final column of Table 1 shows how the predictors are weighted. The W weights that explain which (and to what extent) regions have been used to construct the synthetic control region were Stockholm (.133), Östergötland (.539), Kronoberg (.224), Västerbotten (.014) for the younger age group, and Stockholm (.199), Östergötland (.301), Jönköping (.119), Västra Götaland (.232), Jämtland (.148) for the older age group.

Because the mean squared prediction error (MSPE) in the pre-intervention period is close (or equal) to zero in most panels, we use the post-intervention MSPE to rank the effect sizes across regions. The post-treatment differences between Uppsala and the synthetic controls in Figure 3 show the increases in vaccination shares for 16–17-year-olds (left) and 50–59-year-olds (right). The letters with pre-booked appointments seem to have increased vaccinations in the younger age group, while the increase is less pronounced in the older age group. In the last week of the sample period, week 46, the vaccination rate was 11.7 (84.5 versus 72.8) and 3.6 (93.3 versus 89.6) percentage points higher in the treated region than they would have been in the absence of treatment. The results are very similar if we exclude the five control regions with nudge-like interventions directed at much older age groups as described in Section 2 (see Figure A1 in the Appendix).

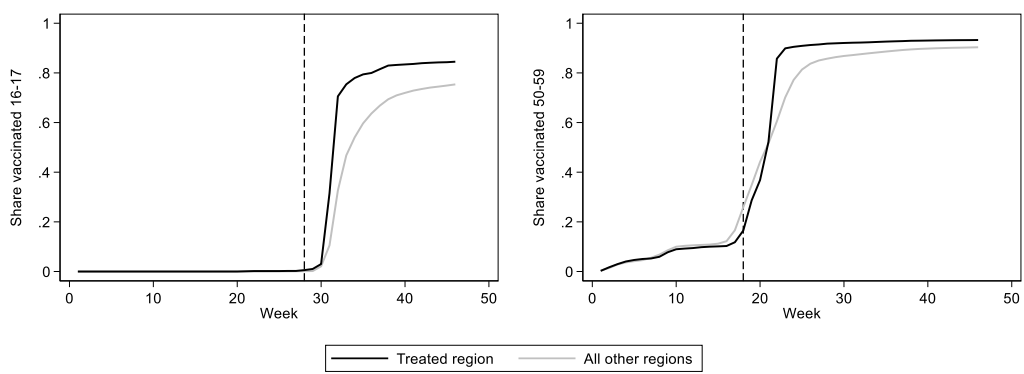

Figure 2. Share of vaccinated 16-17-year-olds (left) and 50-50-year-olds (rights) in the treated region (Uppsala) versus the average of all other regions, by week before and after intervention (vertical line)

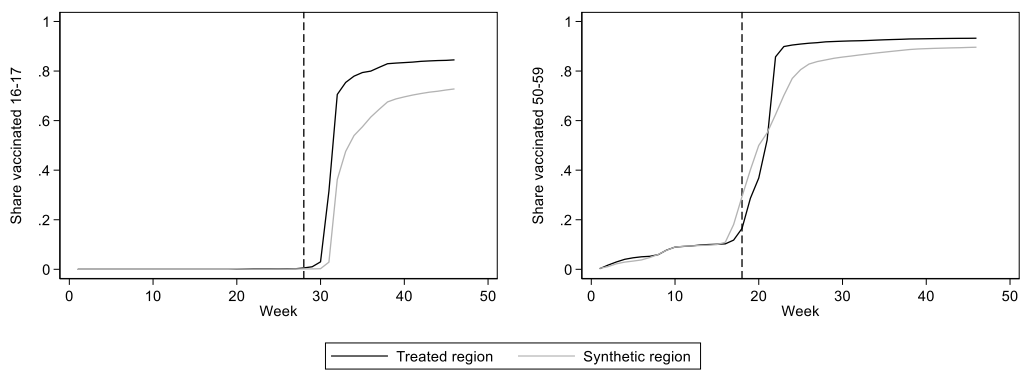

Figure 3. Path plot of the share vaccinated 16–17-year-olds (left) and 50–59-year-olds (right): Treated region (Uppsala) versus the synthetic region

Table 1. Vaccination share predictor means

|  | Uppsala | Synthetic Uppsala | Average of 20 control regions | V |
|---|---|---|---|---|
| Panel A: 16–17 years | | | | |
| Share foreign-born | .189 | .182 | .160 | .361 |
| Share high education | .180 | .157 | .138 | .033 |
| Share of COVID-19 deaths | .001 | .001 | .001 | .010 |
| Share with financial aid | .041 | .042 | .040 | .090 |
| Alcohol addiction per 100k | 54.0 | 69.6 | 89.1 | .044 |
| Share with fast internet | .862 | .868 | .836 | .265 |
| Share with high trust | .752 | .750 | .737 | .076 |
| Share NEETs | .062 | .070 | .078 | .085 |
| Share vaccinated (18–29 y) | .077 | .075 | .074 | .035 |
| Panel B: 50–59 years | | | | |
| Share foreign-born | .189 | .185 | .160 | .037 |
| Share high education | .180 | .161 | .138 | .014 |
| Share of COVID-19 deaths | .001 | .001 | .001 | .075 |
| Share with financial aid | .041 | .036 | .040 | .029 |
| Alcohol addiction per 100k | 54.0 | 72.2 | 89.1 | .030 |
| Share with fast internet | .862 | .867 | .836 | .052 |
| Share with high trust | .752 | .746 | .737 | .072 |
| Share NEETs | .062 | .072 | .078 | .033 |
| Share vaccinated (50–59 y) | .069 | .070 | .077 | .658 |

*Notes:* The period for each predictor is 2020, except for Share vaccinated, which refers to the mean share for all pre-intervention weeks. V is the variable importance weights.

## 4.3 Robustness Tests

We perform a series of tests to assess the robustness of our results. First, we conduct an in-space placebo test and a leave-one-out analysis, as suggested by Abadie *et al.* (2010) and Abadie (2021). More importantly, synthetic control analyses can be sensitive to how one chooses to input the pre-intervention outcomes into the optimization procedure. Following Ferman *et al.* (2020), we also estimate 14 different specifications where we vary how we input the pre-intervention outcomes in the synthetic control optimization method.

For the in-space placebo tests, we compare the effect estimated for the treated region with the effect of placebo interventions iteratively implemented in the non-treated 20 regions. This shows us if the results obtained for the treated region are large compared to the placebo results in the other regions, and it allows for measuring the fraction of regions with results as big as those of the treated region (placebo-based p-values). The results are presented in Figure 4. For the 16–17-year-olds (left panel) no other region has an effect estimate close to the one in the treated region; a ranking of the post-intervention effect sizes across all regions shows that Uppsala has by far the largest estimated effect, with a placebo-based p-value of 1/21=0.048. Overall, the analysis indicates a large and persistent effect of the intervention on 16–17-year-olds. For the 50–59-year-olds (right panel) the effect estimate in the treated region is not that different from the placebo effect estimates. A ranking of the effect sizes in this age group shows that the treated region does not clearly stand out compared to the untreated regions, with a placebo-based p-value of 3/21=0.143 (see Figure A2 in the Appendix). To assess if the two age-specific effect estimates are different from each other, we also perform an analysis where we compute the difference in effect estimates between 16-17-year-olds and 50-59-year-olds by week relative to the start of vaccinations in each age group. Again, we perform the analysis within Uppsala and compare the results to all other regions for placebo-based inference. We find that the difference in effect estimates between the two groups is among the largest in the Uppsala region, with a placebo-based p-value of 2/21=0.095 (Figure A3).

We conduct a leave-one-region-out robustness analysis, as suggested by Abadie (2021). One at a time, we take out each of the four/five regions that contribute to the synthetic control to ensure that the results are not entirely driven by the inclusion of a specific control region. All leave-one-out estimates closely track the findings from the main analysis. Thus, the main findings appear robust (see Figure A4 in the Appendix).

Following Ferman *et al.* (2020), we also estimate 14 different specifications in which we vary how we input the pre-intervention outcomes in the synthetic control optimization procedure and find that the effect size is largest in the treated region in all 14 specifications for 16–17-year-olds. For 50–59-year-olds, we find that the effect size is largest in the treated region in only one of the specifications. This implies that the effect estimate for the younger age group is robust to specification searching. However, for the older age group, the effect estimate is not robust (see Table A2 in the Appendix).

Finally, we assess the vaccination share for other age groups. Since 18–29 and 40–49-year-olds in Uppsala were not sent letters with pre-booked appointments, we do not expect them to have a higher vaccination rate. However, there may be spillovers from the treatment; increased vaccinations in the treated age group may increase vaccinations among friends and relatives in the untreated age group. From the effect plots, we see an indication of an effect for both these age groups, but placebo estimations show that those effects are not only small but also statistically insignificant (see Figures A5 and A6 in the Appendix). In Uppsala, older age groups were sent letters with pre-booked appointments a few weeks before the 50-59-year-olds, the effects in this groups are even less pronounced than in the 50-59-year age group (see Figures A7-A9 in the Appendix).

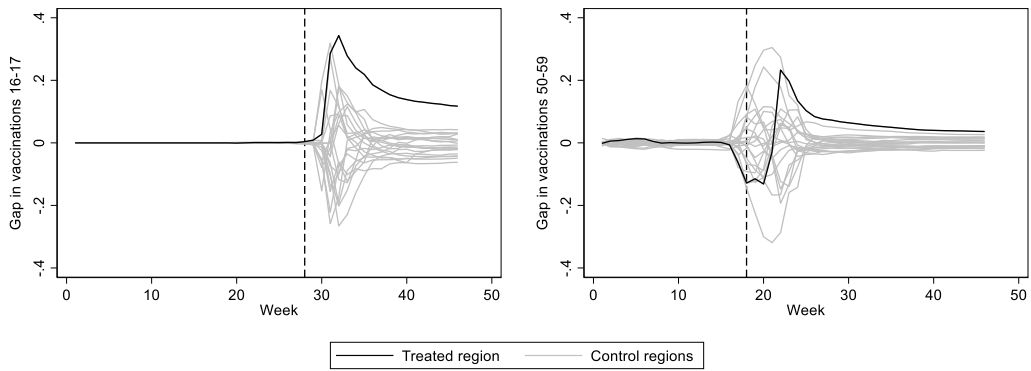

Figure 4. Permutation tests: vaccination share gap in the treated region and placebo gaps for the control regions, 16–17-year-olds (left), 50–59-year-olds (right)

## 5 Discussion and Conclusion

Our results suggest that persons at low risk of adverse events due to Covid-19 are easily nudged into taking the vaccine, but those at higher risk are not. Thus, this paper empirically shows that a default nudge can be successful in impacting decisions if the nudged individuals have sufficiently little intrinsic motivation to prefer one choice over another. Our main innovation is to use two groups, each subjected to the same intervention and each with its own control group.

Our setting is well-suited for measuring the impact of the nudge because the vaccine was rolled out quickly and the letters with pre-booked appointments were sent to both the quinquagenarians and the teenagers at roughly similar times between May and July of 2021, a time of the year that is off-season for Covid-19 in Sweden, so the two groups made their vaccination decisions in a similar pathogenic environment. In addition, the age-stratified mortality and morbidity of Covid-19 create a natural differentiation in intrinsic motivation for the two groups to get vaccinated. Thus, natural factors and the auspicious timing of the vaccine roll-out combine to create a clean setting for the impact of nudges to be estimated. In addition to shedding light on the effects of nudges more generally, our findings directly relate to the empirical literature on the effects of nudges and incentives to increase Covid-19 vaccinations (Barber & West, 2021; Bonander *et al.*, 2022; Dai *et al.*, 2021; Campos-Mercade et al., 2021).

Our interpretation of the findings in this paper relies on the assumption that pre-booked vaccinations influence behaviors via nudging. However, one factor unrelated to nudging that we cannot eliminate is the impact of who makes the bookings. If individuals make their own arrangements, popular times may fill up quickly and people may be reluctant to choose less suitable times. Time off work for Covid-19 vaccinations is widely accepted in Sweden, but it probably looks better if a worker leaves in the middle of the day because he has been called to a vaccination appointment than if he leaves because he has chosen it. Thus, the very small effect, if any, we observe among 50-59-year-olds may be due to a coordination factor unrelated to nudging.

Nudges are known to have highly variable and context-specific effects (Szaszi *et al.*, 2018, Löfgren and Nordblom, 2020, de Ridder *et al.*, 2022). A natural factor that moderates their impact is intrinsic motivation. In light of the stark contrast in the effect of default nudges on vaccine take-up that we find, we believe that nudge theory would benefit from more efforts to highlight the effect moderation in heterogeneous groups of nudged individuals, to ascertain when nudges are powerful influences on action and when they are not.

# Appendix

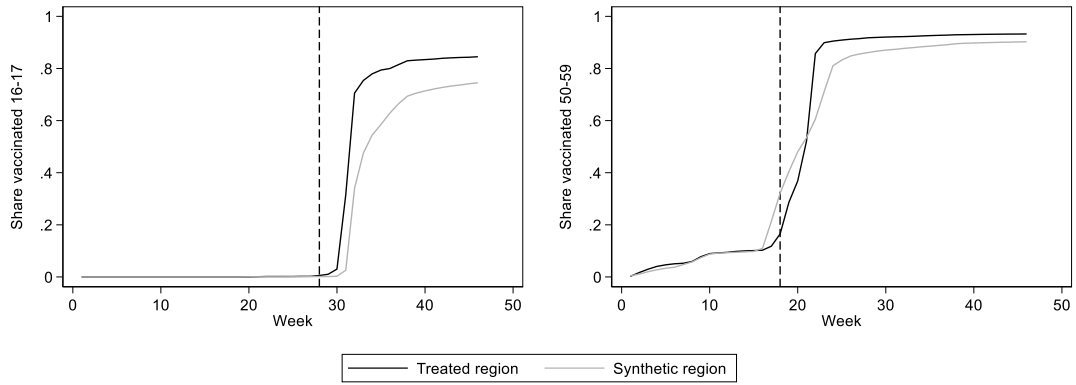

Figure A1. Path plot of the share vaccinated 16–17-year-olds (left) and 50–59-year-olds (right): Treated region (Uppsala) versus the synthetic region, including only 15 regions in the donor pool

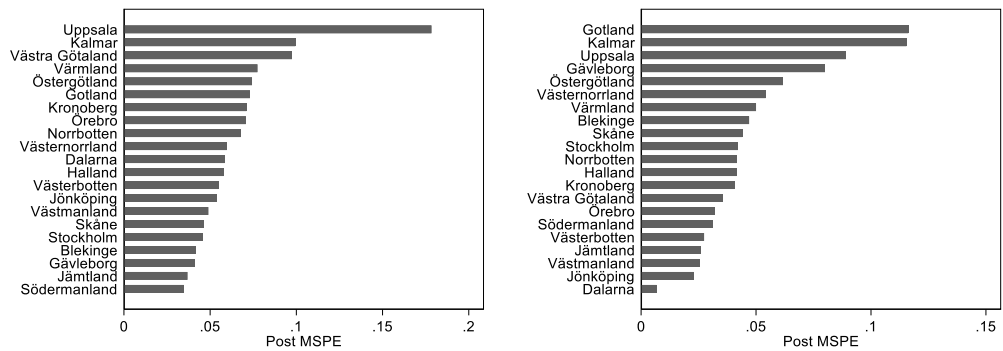

Figure A2. Post-intervention MSPE for 16-17-year-olds (left) and the post-intervention MSPE for 50-59-year-olds (right) for the treated region (Uppsala) and control regions

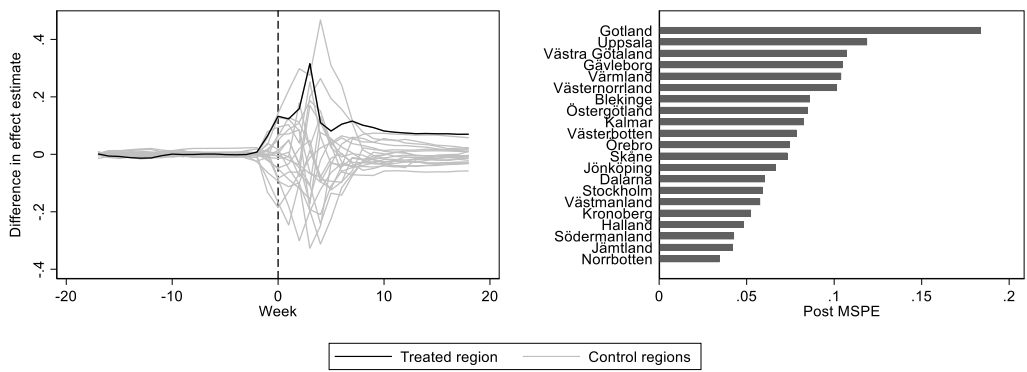

Figure A3. Difference in effect estimates between 16-17-year-olds and 50-59-year-olds by week relative to the start of vaccinations in each age group (left) and post-intervention MSPE for the difference in effect between the two age groups (right) in Uppsala and control regions

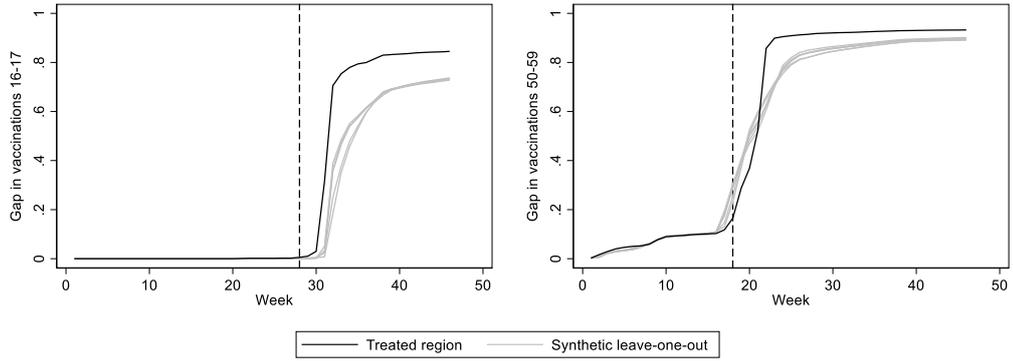

Figure A4. Leave-one-out distribution of the synthetic Uppsala for 16-17-year-olds (left panel) and 50-59-year-olds (right panel)

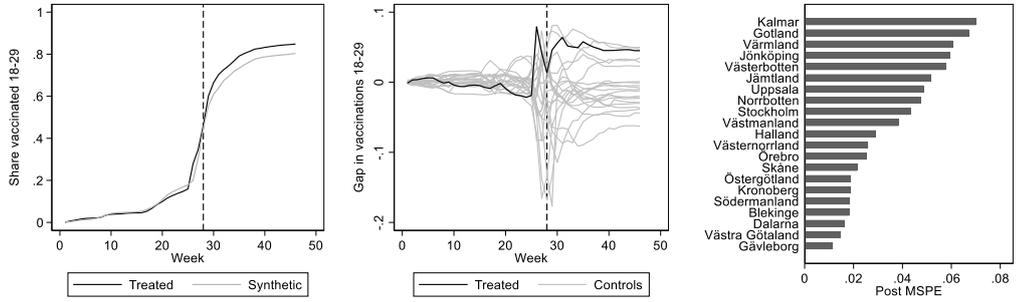

Figure A5. Effect (left), placebo (middle), and post-intervention effect size (right) vaccinations among 18-29-year-olds (untreated group)

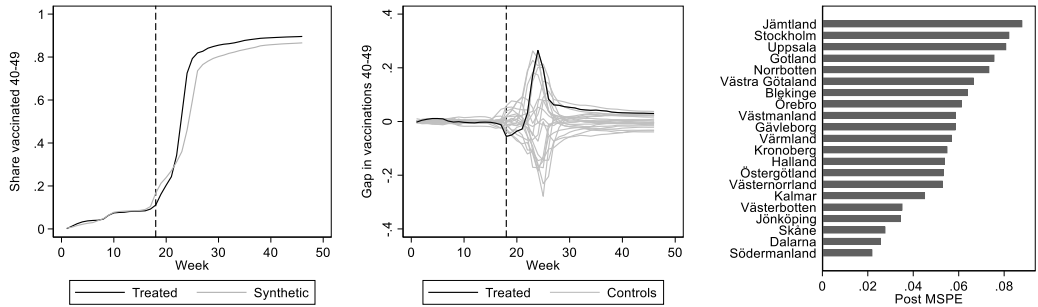

Figure A6. Effect (left), placebo (middle), and post-intervention effect size (right) vaccinations among 40-49-year-olds (untreated group)

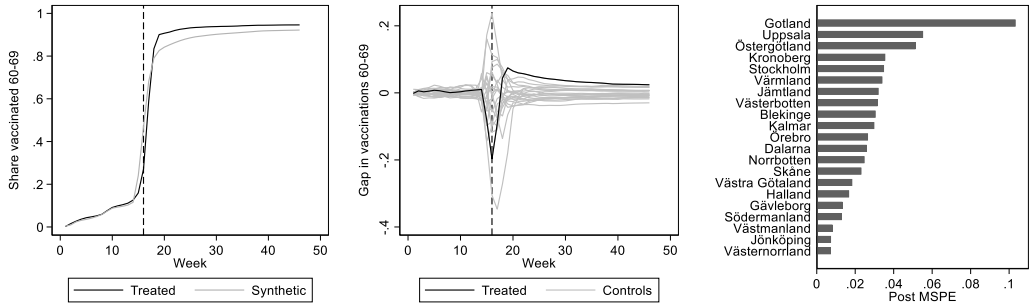

Figure A7. Effect (left), placebo (middle), and post-intervention effect size (right) vaccinations among 60-69-year-olds

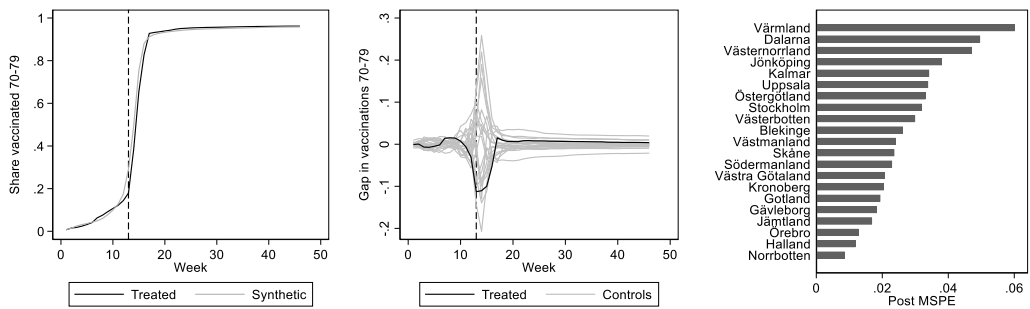

Figure A8. Effect (left), placebo (middle), and post-intervention effect size (right) vaccinations among 70-79-year-olds

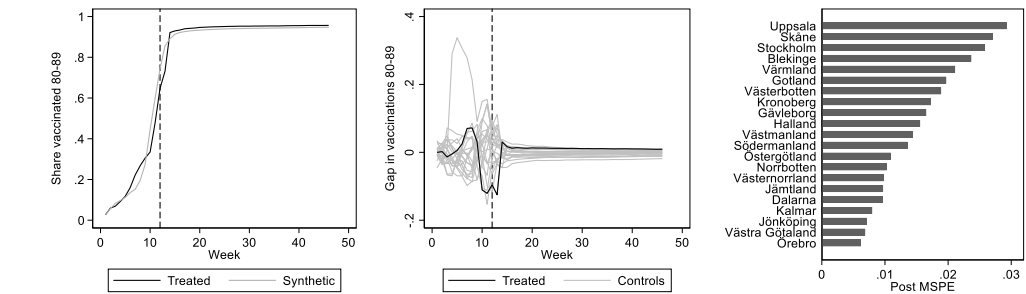

Figure A9. Effect (left), placebo (middle), and post-intervention effect size (right) vaccinations among 80-89-year-olds

Table A1. Specifications searching

| Panel A: 16–17 years | | | | | | | |
|---|---|---|---|---|---|---|---|
| Specification | (1a) | (1b) | (2a) | (2b) | (3a) | (3b) | (4a) |
| p-value | 0.048 | 0.048 | 0.048 | 0.048 | 0.048 | 0.048 | 0.048 |
| Specification | (4b) | (5a) | (5b) | (6a) | (6b) | (7a) | (7b) |
| p-value | 0.048 | 0.048 | 0.048 | 0.048 | 0.048 | 0.048 | 0.048 |
| Panel B: 50–59 years | | | | | | | |
| Specification | (1a) | (1b) | (2a) | (2b) | (3a) | (3b) | (4a) |
| p-value | 0.143 | 0.095 | 0.190 | 0.190 | 0.143 | 0.095 | 0.238 |
| Specification | (4b) | (5a) | (5b) | (6a) | (6b) | (7a) | (7b) |
| p-value | 0.238 | 0.143 | 0.190 | 0.048 | 0.143 | 0.238 | 0.095 |

*Notes:* Specifications refer to: (1) all pre-treatment vaccination shares (i.e., including each weekly pre-treatment outcome as a covariate), (2) the first three-fourths of the values, (3) the first half of the values, (4) odd pre-treatment values, (5) even pre-treatment values, (6) pre-treatment mean, and (7) three values. Specifications ending with b include all additional eight covariates, while specifications ending with a, include no additional covariates. Since our data have (almost) no variation in the outcome variable in any region before the vaccination rollout for 16–17-year-olds, we use the vaccination share among 18-29-year-olds as the main variable to match on in Panel A.